\documentclass[%
 reprint,
superscriptaddress,
 amsmath,amssymb,
 aps,
 longbibliography,
]{revtex4-1}

\usepackage{graphicx}
\usepackage{dcolumn}
\usepackage{bm}
\usepackage{xr}
\usepackage[colorlinks=true, allcolors=blue]{hyperref}
\usepackage{color}
\usepackage{braket}
\usepackage{multirow}

\newcolumntype{P}[1]{>{\centering\arraybackslash}p{#1}}

\begin{document}

\preprint{APS/123-QED}

\title{Control of magnetic interactions between surface adatoms via orbital repopulation }

\author{Danis I. Badrtdinov}
\affiliation{\mbox{Theoretical Physics and Applied Mathematics Department, Ural Federal University, 620002 Ekaterinburg, Russia}}

\author{Alexander N. Rudenko}
\email{a.rudenko@science.ru.nl}
\affiliation{School of Physics and Technology, Wuhan University, Wuhan 430072, China }
\affiliation{\mbox{Theoretical Physics and Applied Mathematics Department, Ural Federal University, 620002 Ekaterinburg, Russia}}

\author{Mikhail I. Katsnelson}
\affiliation{\mbox{Institute for Molecules and Materials, Radboud University, Heijendaalseweg 135, 6525 AJ Nijmegen, Netherlands}}
\affiliation{\mbox{Theoretical Physics and Applied Mathematics Department, Ural Federal University, 620002 Ekaterinburg, Russia}}

\author{Vladimir V. Mazurenko}
\email{vmazurenko2011@gmail.com}
\affiliation{\mbox{Theoretical Physics and Applied Mathematics Department, Ural Federal University, 620002 Ekaterinburg, Russia}}

\date{\today}

\begin{abstract}
We propose a reversible mechanism for switching Heisenberg-type exchange interactions between deposited transition metal adatoms from ferromagnetic to antiferromagnetic. Using first-principles calculations, we show that this mechanism can be realized for cobalt atoms on the surface of black phosphorus by making use of electrically-controlled orbital repopulation, as recently demonstrated by scanning probe techniques [Nat. Commun. {\bf 9}, 3904 (2018)]. 
We find that field-induced repopulation not only affects the spin state, but also causes considerable modification of exchange interaction between adatoms, including its sign. Our model analysis demonstrates that variable adatom-substrate hybridization is a key factor responsible for this modification. We perform quantum simulations of inelastic tunneling characteristics and discuss possible ways to verify the proposed mechanism experimentally.
\end{abstract}
\maketitle


\section{\label{sec:introduction}Introduction}
Controllable manipulation of interactions between magnetic atoms is one of the primary goals in spintronics. For instance, it can be achieved using high-frequency electromagnetic field~\cite{Rasing2010, Radu2011, Mikhaylovskiy2015}, which opens the possibility for switching of magnetic interactions in collinear magnets~\cite{Li2013} or in more complex spin textures like skyrmions~\cite{Stepanov2017,Titov2017}.  Ultracold atoms in optical lattices provide another possibility for engineering the exchange interactions in magnetic systems~\cite{Anderlini2007, Trotzky2008}. Despite vast perspective, practical application of these methods is limited. Alternative way is offered by scanning tunneling microscopy (STM) techniques allowing for selective manipulation of magnetic interactions at the level of individual atoms deposited on surfaces~\cite{MnCuN,CoPt,Khajetoorians2012,Khajetoorians2019}.  Progress in this field is especially promising for applications, as it paves the way to create logic nanodevices \cite{Kha}, magnetic storage \cite{Loth}, sensors \cite{Loth1}, and other elements required for further miniaturization of computing units.

\begin{figure}[!tbp]
\includegraphics[width=0.51\textwidth]{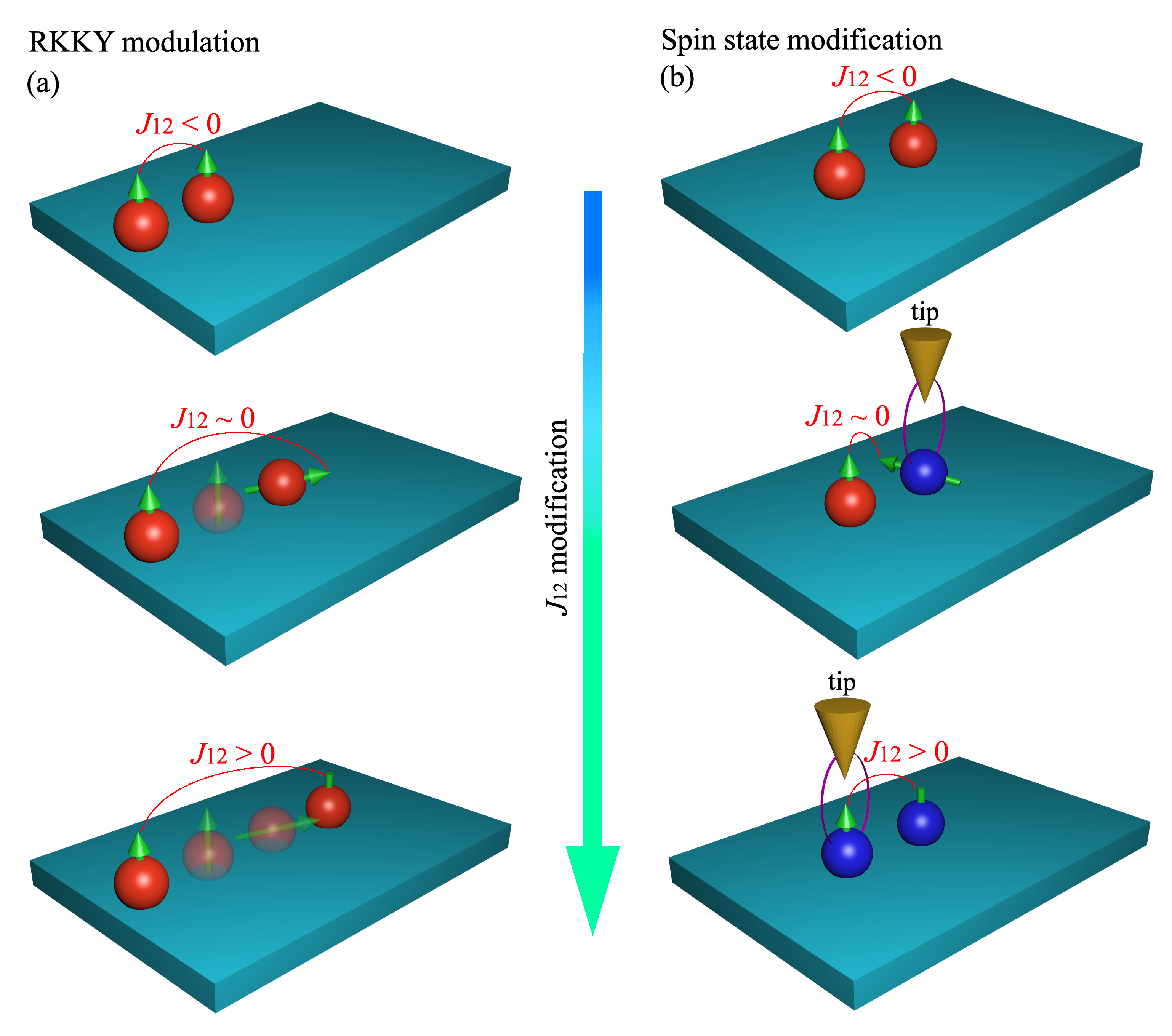}
\caption{Two scenario realizing controllable switching of magnetic interactions between adatoms on a surface. (a) Conventional distance-dependent RKKY mechanism based on the control of magnetic coupling by means of adatom relocation. (b) Switching mechanism that is proposed in this work, based on the selective spin state modification of adatoms using STM tip.}
\label{fig:Model}
\end{figure}

At the same time, experimental approaches to manipulate magnetic interactions by an STM tip are rather limited. For instance, substrate-mediated interaction of the Ruderman-Kittel-Kasuya-Yoshida (RKKY) type between magnetic adatoms on metallic surfaces can be tuned by tailoring the interatomic distance with the STM tip, as it was demonstrated, for instance, for magnetic adatoms on Cu(111)~\cite{Khajetoorians2012} and Pt(111)~\cite{Zhou2010, CoPt} surfaces. As a result, one can switch between ferromagnetic (FM), zero, and weak antiferromagnetic (AFM) Heisenberg exchange interactions at various distances, as it is schematically shown in Fig.~\ref{fig:Model}(a). 
On the other hand, the interactions between adatoms on insulating surfaces originate from superexchange, being typically of AFM-type~\cite{Rudenko2009, MnCuN}, and decaying very fast as interatomic distance increases.

In this work, we propose another mechanism for switching of magnetic interactions that does not require relocation of adatoms. Our idea is based on field-assisted orbital repopulation of adatoms by means of STM tip, leading to a modification of adatom spin state. Recently, such a possibility was experimentally demonstrated in Ref.~\onlinecite{Khajetoorians} for a system of individual cobalt (Co) adatoms deposited on black phosphorus (BP) surface. It was shown that orbital population of the $d$-shell and, consequently, magnetic moment of individual Co atoms can be effectively manipulated by applying a bias voltage between STM tip and the surface. As we will show below, similar manipulations with a Co dimer can results in the reversible switching of magnetic interactions from FM to AFM, as it is sketched in Fig.~\ref{fig:Model}(b).

Our interest to BP substrate is not only triggered by its practical relevance in virtue of attractive semiconducting properties, thickness-dependent band gap~\cite{Li2017}, high carrier mobility~\cite{Rudenko2016, Li2014}, and intrinsic $p$-doping \cite{Kiraly2017}. Being a highly anisotropic electron system, BP hosts unusual properties of fundamental importance, related mainly to dielectric screening and optical response \cite{Serrano2016,Prishchenko_2017,Kiraly2019}.
Recent numerical simulations revealed that RKKY interaction between magnetic impurities on doped monolayer BP (phosphorene) is highly anisotropic, and largely determined by the orientation of impurities~\cite{RKKY_Zare, RKKY_Duan}.
Similar conclusion has been made explicitly for Co/BP system, previously analyzed from first principles in the presence of doping~\cite{Catro_Neto} and mechanical strain~\cite{Cai2017}. Despite existing interest to the Co/BP system, no attention has been paid so far to the problem of interaction engineering.

The rest of the paper is organized as follows. Sec.~\ref{sec:Results} presents the results of first-principles calculations of exchange interactions between for two cobalt atoms on BP. In Sec.~\ref{sec:model}, we present an effective model, which qualitatively describes the mechanism of FM-AFM switching upon changing of the spin states of adatoms. In Sec.~\ref{sec:discuss}, we discuss possible ways for experimental detection of the switching observed. Sec.~\ref{sec:conc} concludes the paper.

\section{\label{sec:Results}First-principles calculations}
\subsection{Method and computational details}
The main focus of our study is to obtain isotropic exchange interactions between cobalt adatoms deposited on monolayer BP. To this end, we consider the Heisenberg Hamiltonian in the following form:
\begin{eqnarray}
\mathcal {\hat{H}}_{12} = J_{12}\hat{\mathbf{S}}_1\hat{\mathbf{S}}_2,
\label{eq:Ham}
\end{eqnarray} 
where spins of each cobalt atoms are represented by operators $\hat{\mathbf{S}}_1$ and $\hat{\mathbf{S}}_2$, which can be different depending on the orbital configuration of cobalt atom \cite{Khajetoorians}, resulting in either high-spin or low-spin state.  $J_{12}$ is the exchange interaction parameter, which we determine from first-principles calculations. To this end, we consider the energies of collinear FM and AFM spin configurations, yielding $J_{12} =(E_{\mathrm{FM}} - E_{\mathrm{AFM}})/{2S_1 S_2}$, where spin $S$ of each atom takes the values $S = 1/2$ or $S = 1$ for the low- and high-spin state, respectively.  In Eq.~(\ref{eq:Ham}), we do not consider anisotropic terms. As it was shown in Ref.~\onlinecite{Khajetoorians}, the magnetic anisotropy energy in cobalt on BP is weak, with magnitude of $\sim$0.1 meV. On this basis we also expect that the inter-site anisotropic interactions (such as Dzyaloshinskii-Moriya interaction) are also weak in comparison with the isotropic exchange interaction providing the main contribution to the magnetic energy of the system in question. 

To calculate the energies of collinear magnetic configurations, we use density functional theory (DFT) electronic structure calculations performed within the generalized gradient approximation in the parametrization of Perdew-Burke-Ernzerhof (GGA-PBE)~\cite{PBE}. We employ the projected augmented wave method \cite{Blochl} as implemented in \emph{Vienna ab-initio simulation package} ({\sc vasp})~\cite{Kresse,Furthmuller}. 

In the calculations, we set the plane-wave energy cutoff to 300 eV and the energy convergence criteria to 10$^{-6}$ eV, which was checked to be sufficient to achieve numerical convergence of exchange interactions. The surface was modelled in the slab geometry by a single BP layer using experimental structure~\cite{Exp_BP} with the lattice constants $a$ = 3.313 \AA \, and $b$ = 4.374 \AA.  In order to avoid spurious interactions between the cobalt atoms at large distances, the following supercells were used: $5a \times 8b$ (160 atoms) and $8a \times 4b$ (128 atoms) for the cobalt dimer oriented along the armchair and zigzag crystallographic directions, respectively. The corresponding Brillouin zone was sampled by $\Gamma$-centered (4$\times$3) and (3$\times$4)  {\bf k}-point meshes, respectively. A vacuum space of 18~\AA \, was introduced between the unit cell images in the direction normal to the surface.  The position of cobalt atoms was optimized in each case considered until the residual force were less than 0.005 eV/\AA.

To simulate low- and high-spin states of cobalt adatoms, we consider on-site Coulomb repulsion in the $d$-shell, which is known as one of the key factors governing the orbital configuration of adsorbed transition metal atoms \cite{Khajetoorians,Rudenko2012,Sessi2014}. Specifically, we apply the Hubbard-$U$ correction at the mean-field level using a simplified rotatonally invariant version ~\cite{Dudarev} of the DFT+$U$~\cite{Anisimov} scheme. In these calculations, an effective interaction  $\tilde{U}= U -J_H$ is assumed ($J_H$ is the Hund's rule coupling), playing the role of a phenomenological parameter related to the Coulomb repulsion. 
Previously, it was found that the variation of $\tilde{U}$ parameter in the range 0 -- 3 eV stabilizes the low-spin state of cobalt on BP, while higher values of $\tilde{U}$ makes the high-spin state more favorable~\cite{Khajetoorians}. For this reason, here we use $\tilde{U}= 0$ eV and 4 eV to simulate low- and high-spin states, respectively.

In what follows, we consider two inequivalent orientations of the cobalt dimer on BP, which correspond to the zigzag ($X$) and armchair ($Y$) directions. In each case, we choose the hollow position of cobalt adatoms, which is the most energetically favorable adsorption site for both spin states considered \cite{Khajetoorians}. The calculations are then performed for various interatomic distances, including optimization of the adatom positions.

\subsection{Short-range interactions}
We first consider short-range interaction between closest cobalt adatoms on BP. Table~\ref{tab:short_range} gives the resulting magnetic moments and isotropic exchange interactions for three different combinations of spin states, namely, high-high (HH), high-low (HL), and low-low (LL).  For the low-spin configuration, the total magnetic moment per unit cell containing two cobalt atoms is equal to 2~$\mu_B$, which means that the spin state of cobalt atom can be associated with spin $S$ = 1/2 (1~$\mu_B$). The deviation of single adatom moment (0.81~$\mu_B$) from the 1~$\mu_B$ can be attributed to hybridization of cobalt and phosphorus states.  This indicates that the magnetic moment is not entirely localized on cobalt, but is spread out over neighboring phosphorus atoms. On the other hand, the total magnetic moment on adatoms in the high-spin state is slightly higher than for two spins $S = 1$, yet individual adatom moments are very close to 2 $\mu_B$.  In agreement with previous studies~\cite{Khajetoorians}, the high-spin state is characterized by a larger adatom-substrate distance $d$ compared to the low-spin state (see Table~\ref{tab:short_range}). This suggests different hybridization strength for the two states, which can lead to the realization of diverse exchange interaction mechanisms between adatoms in different spin states.  
To assess the relative hybridization strength, we estimate a correlation function $\langle D_{Co}(\epsilon) D_{P}(\epsilon) \rangle$, which quantifies the overlap of non-spin-polarized density of states (DOS) $D_{Co}(\epsilon)$ and $D_{P}(\epsilon)$ projected onto cobalt and phosphorus states, respectively. Specifically, we define $\langle ... \rangle = \frac{1}{\braket{D}} \int_{-\infty}^{E_F} (...) d \epsilon$ with $\braket{D}$ being the average density of occupied states. For single cobalt adatom, we obtain 6.7 and 5.0 using the structure of low-spin and high-spin configuration, respectively. This indicates that the low-spin state of cobalt atom is more hybridized with the substrate.

\begin {table}[!btp]
\centering
\caption [Bset]{Summary of first-principles calculations obtained in this work for nearest-neighbor cobalt atoms on BP. The combination of different spin states is denoted as high-high (HH), high-low (HL), and low-low (LL). } 
\begin{ruledtabular}
\setlength{\extrarowheight}{3pt}
\label{tab:short_range}
\begin {tabular}{c|P{1cm}|P{2cm}|P{1cm}}
 Configuration       &  HH    & HL & LL  \\
 \hline
Exchange interaction (meV)  & 13.3 &  4.8  &   -51.2   \\
Co-Co distance $R$ (\AA) & 4.34 & 4.47 &  4.37   \\
Co-BP distance $d$ (\AA) & 1.37  & 1.34/1.00 &  1.01   \\
Total magnetic moment ($\mu_B$) & 4.20 & 3.00   &   2.00 \\
Adatoms moments ($\mu_B$) & 1.91  & 1.91/0.55  &  0.81  \\
\end {tabular}
\end{ruledtabular}
\end {table}
 
\begin{figure}[!tbp]
\includegraphics[width=0.48\textwidth]{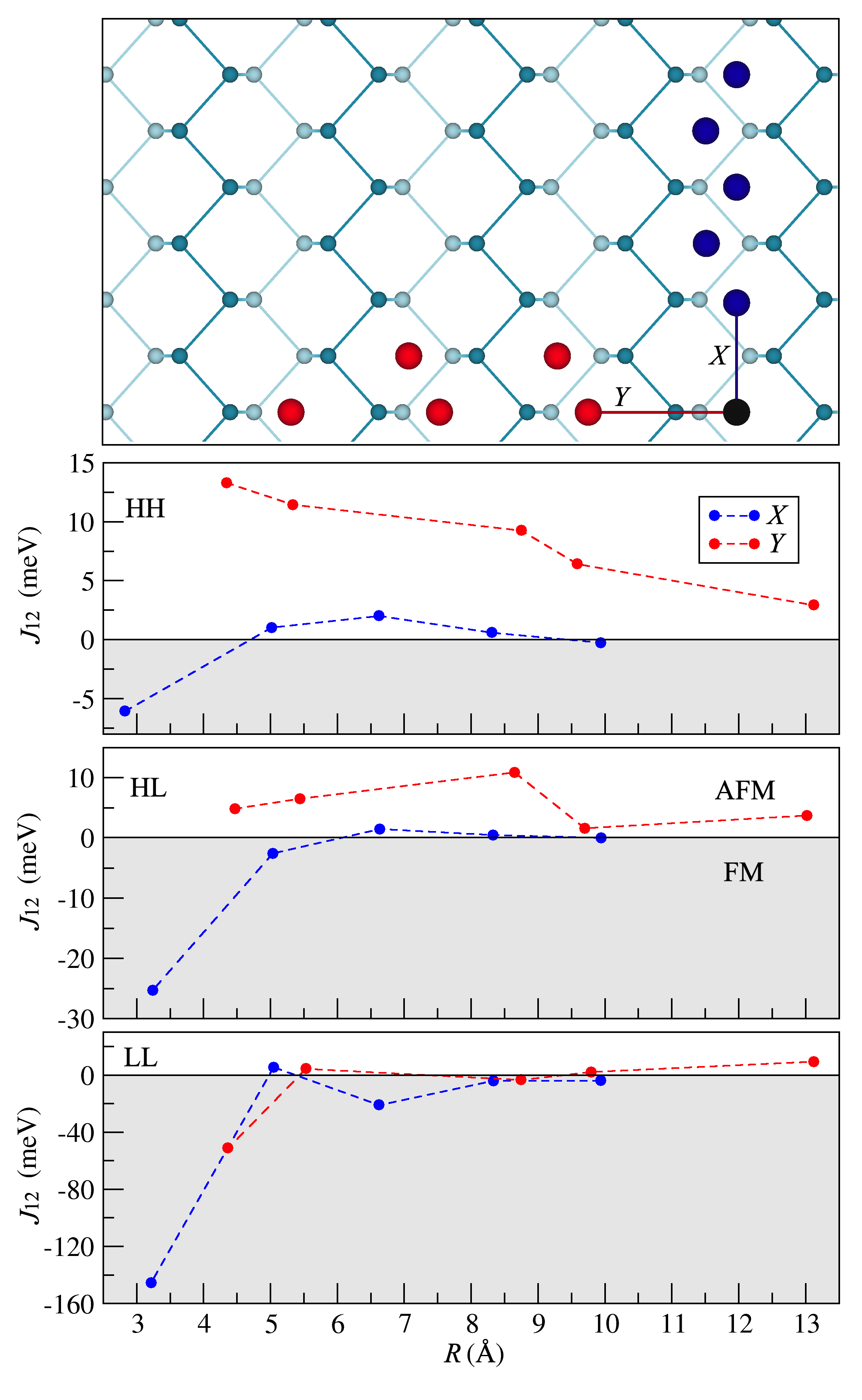}
\caption{(Top) Schematic crystal structure of phosphorene with cobalt atoms on top. Small and big spheres correspond to phosphorus and cobalt atoms, respectively. Black sphere denotes a reference cobalt atom, whereas blue and red spheres correspond to different positions of the second cobalt atom forming a dimer along the zigzag ($X$) and armchair ($Y$) direction, respectively. 
(Bottom) Calculated exchange interactions $J_{12}$ between cobalt adatoms shown as a function of interatomic distance for different spin states of the adatom: HH, HL, and LL. White and gray background show AFM and FM coupling, respectively. }
\label{fig:Long_range}
\end{figure}

The magnetic interactions in the pair of cobalt atoms deposited along the armchair direction is indeed very sensitive to the adatom spin state. From Table \ref{tab:short_range}, one can see that FM and AFM coupling is realized for LL and HH configurations, respectively.  Furthermore, for a mixed  spin combination (HL), we observe a considerably weaker interaction between the adatoms.   

On the other hand, such behavior is not observed for atoms in the nearest-neighbor zigzag positions -- for all three spin configurations $J_{12}$ is ferromagnetic. This can be related to the following factors: (i) Along the zigzag direction, cobalt adatoms are placed much closer to each other, enhancing the FM direct exchange coupling~\cite{Badrtdinov2016, Badrtdinov2018}; (ii) The Co-P-Co angle for zigzag HH and LL configurations equals to 74.5$^\circ$ and 96.1$^\circ$, whereas for armchair it is 114.2$^\circ$ and 129.5$^\circ$, respectively. According to the Goodenogh-Kanamori rules~\cite{Goodenough1958, Kanamori1959}, metal-ligand-metal bond angle plays a crucial role in the superexchange coupling. While AFM exchange is favored in case of a 180$^\circ$ bond, the exchange tends to FM if a 90$^{\circ}$ bond is realized. The latter is consistent with the FM interaction observed for zigzag alignment of cobalt adatoms. (iii)  Finally, black phosphorus substrate, being a strongly anisotropic system, is expected to provide an indirect contribution to the exchange depending on the direction of adatom alignment~\cite{RKKY_Zare, RKKY_Duan}.

 In order to check the robustness of the obtained results with respect to the Hubbard $U$ parameter, we calculate exchange interactions for the case of armchair direction using  $\tilde{U}= 1$ eV and 5 eV for  low- and high-spin states, respectively, which correspond to a smaller dielectric screening. We find that it leads to a reduction of exchange interaction $J_{12}$ from 13.3 eV to 12.3 eV for HH and from $-51.2$ eV to $-28.0$ eV for LL spin configuration. This can be explained by a stronger electronic localization induced by larger Coulomb interaction $\tilde{U}$. Nevertheless, the sign of the interaction remains unchanged with respect to a small variation of $\tilde{U}$, meaning that the spin state of a cobalt atom plays the decisive role in determining the magnetic interaction.

The complexity of the electronic structure prevents us from providing an immediate microscopic interpretation of the observed behavior. To address this problem, in Sec.~\ref{sec:model}, we construct an effective model, which reveals that hybridization effects play a key role in the formation of magnetic coupling, making either FM of AFM alignment favorable in the cobalt dimer.

\subsection{Long-range interactions}
Having discussed the alternative character of short-range interactions realized between cobalt adatoms on phosphorene, we now analyze their long-range behavior. The main results are shown in  Fig.~\ref{fig:Long_range}, whereas detailed information about the magnetic moments and exchange couplings is summarized in Appendix~\ref{app:ISO}.

The observed behavior is essentially different for different adatom orientations. While the exchange interactions $J_{12}$ exhibit similar behavior for different spin states along the zigzag direction, the interaction between adatoms oriented along the armchair direction depends strongly on the spin state. In the former case (zigzag), the interactions are reminiscent of oscillations, rapidly decaying with distance. Similar behavior was reported for other low-dimensional magnetic systems on insulating substrates \cite{Rudenko2009}. In the armchair case, some variations of $J_{12}$ can be still seen for LL and HL spin configurations, but they almost disappear in the HH case, displaying slow and virtually monotonic decay of AFM $J_{12}$ as a function of distance.

Previous numerical studies \cite{RKKY_Zare, RKKY_Duan} revealed that in case of charge-doped phosphorene, magnetic impurities can interact via Ruderman-Kittel-Kasuya-Yoshida (RKKY) mechanism~\cite{RKKY1,RKKY2,RKKY3}, demonstrating oscillation of exchange couplings in the form $J_{12} (R) \sim \sin (2k_F R)/(2k_F R)^2$, where $k_F$ is the Fermi wave vector. Moreover, periodicity and amplitude of such oscillations depend on the alignment of adatoms (zigzag or armchair),  which results from the electronic structure anisotropy of black phosphorus substrate.  In our calculations we also observe such anisotropy on the level of magnetic interactions.   However, RKKY mechanism cannot be directly applied to the present case because we consider undoped phosphorene, implying the absence of the Fermi surface. On the other hand, it is known that in semiconductors with a bandgap, another indirect mechanism of magnetic coupling is realized, namely the Bloembergen-Rowland (BR) interaction~\cite{BR_interaction}. Contrary to the RKKY mechanism, the BR interactions decay exponentially~\cite{Zhu2011, semiconductors}. We conclude, therefore, that the interactions between cobalt adatoms on BP are of more complex origin, involving a combination of superexchange and indirect mechanisms.

\section{\label{sec:model}Model analysis}
In the previous section, we have demonstrated the possibility to modify the sign of exchange interactions between cobalt adatoms by means of orbital repopulation.
In this section, we construct an effective model, which qualitatively describes the mechanism of FM-AFM switching between the nearest adatoms. 

\begin{figure}[!tbp]
\includegraphics[width=0.48\textwidth]{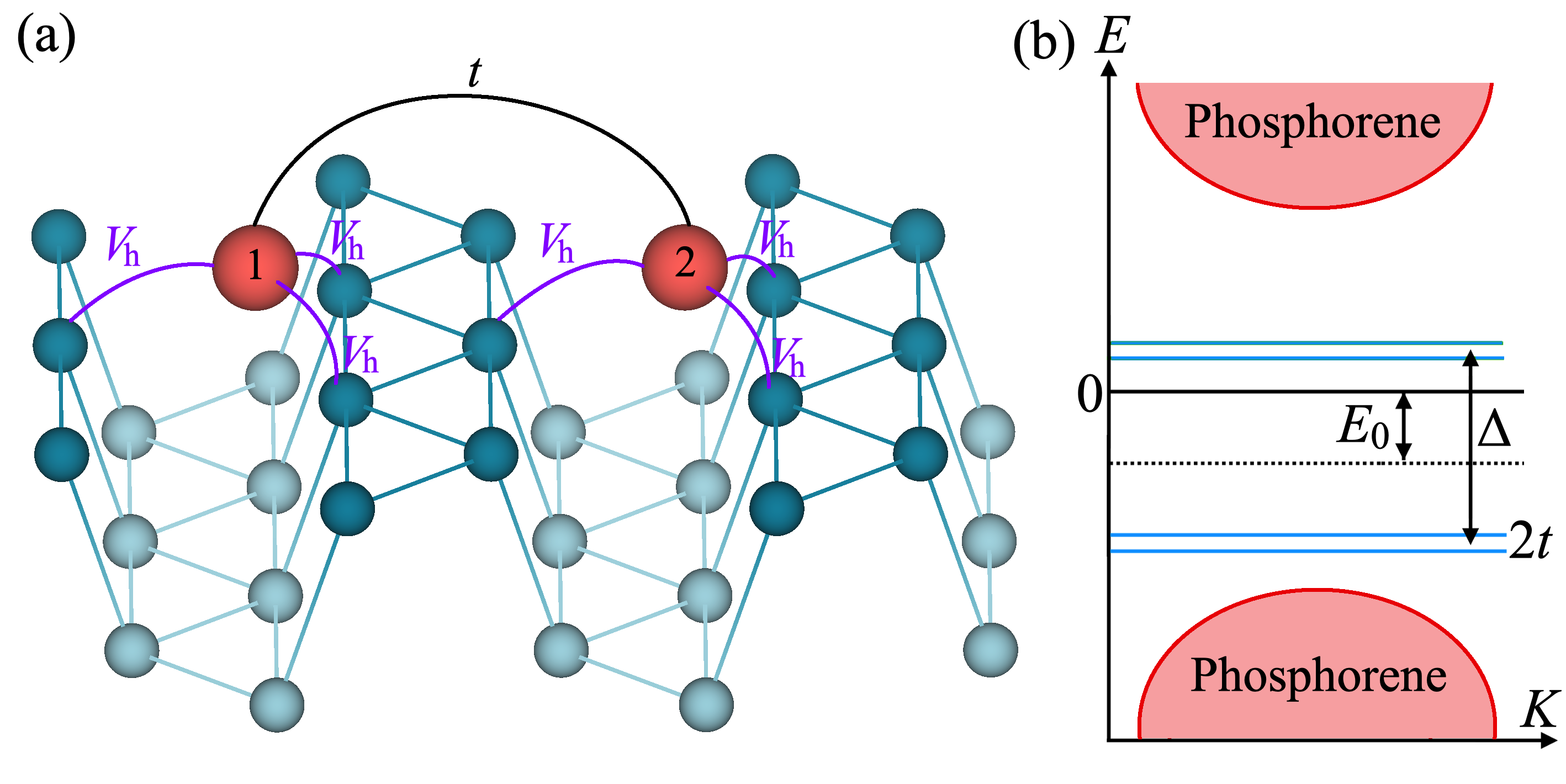}
\caption{ (a) Model of two impurities deposited on the monolayer of BP surface.  Interaction with the substrate is controlled by a hybridization parameter $V$, while direct interaction is realized via $t$ hopping term.  (b) Schematic representation of the band structure of the system in question. Parameter $\Delta$ controls spin splitting, while $E_{0}$ stands for a shift of impurity bands relative to pristine phosphorene states.  Direct hopping parameter leads to energy difference $2t$ between states, belonging two different impurities. The hybridization term $V_h$ introduces finite bandwidth of impurity levels and modifies phosphorene band structure.}
\label{fig:BANDS}
\end{figure}

For this purpose, we consider a model of two impurities in the presence of substrate states. The model is inspired by Ref.~\onlinecite{Anderson_model}, where interaction between two magnetic impurities were examined including the bath of free electron states. Here, we explicitly include BP states in the model, which is schematically shown in Fig.~\ref{fig:BANDS}(a). Each impurity is represented by one half-filled orbital, hybridizing with nearest phosphorus atoms, whereas BP is described within a conventional tight-binding model \cite{Rudenko2015}. It is assumed that the effect of orbital repopulation of cobalt atoms is only captured by the effect of adatom-substrate hybridization. 
The Hamiltonian of such a model takes following form:
\begin{eqnarray}
\mathcal {\hat{H}}= \mathcal {\hat{H}}_{p} + \mathcal {\hat{H}}_{d} + \mathcal {\hat{H}}_{pd},
\label{eq:Ham_model}
\end{eqnarray} 
where $\mathcal {\hat{H}}_{p}$ is the Hamiltonian of pristine BP monolayer
\begin{eqnarray}
\mathcal {\hat{H}}_{p}  =  \sum_{\sigma, i \neq j} t^{||}_{ij} \hat{c}^{\dagger}_{i \sigma} \hat{c}_{j \sigma} +  \varepsilon_0  \sum_{\sigma, i} \hat{c}^{\dagger}_{i \sigma} \hat{c}_{i \sigma},
\label{eq:Ham_model_1}
\end{eqnarray} 
defined in terms of the hopping integrals $t^{||}_{ij}$ taken from Ref.~\onlinecite{Rudenko2015}. Here, the parameter $\varepsilon_0$ ensures that the center of a band gap corresponds to zero energy. Phosphorene was modelled in real space with periodic boundary conditions. For this purpose, we used a supercell containing 48 atoms ($4a \times 3b$), which ensures the absence of interactions between cobalt adatoms in neighboring cells. 

In Eq.~(\ref{eq:Ham_model}), $\mathcal {\hat{H}}_{d}$ is the impurity Hamiltonian, which has the form
\begin{equation}
\begin{aligned}
\mathcal {\hat{H}}_{d}  =    \sum_{\sigma} (E_{0}  - \sigma \Delta) ( \hat{d}^{\dagger}_{1 \sigma} \hat{d}_{1 \sigma} + \hat{d}^{\dagger}_{2 \sigma} \hat{d}_{2 \sigma})   + \\ 
+ t  \sum_{\sigma}  ( \hat{d}^{\dagger}_{1 \sigma} \hat{d}_{2 \sigma} + \hat{d}^{\dagger}_{2 \sigma} \hat{d}_{1 \sigma}).
\end{aligned}
\label{eq:Ham_model_2}
\end{equation} 
The first term here describes the energy $E_{0}$ of impurity states relative to the center of a band gap, while $\Delta$ is the spin splitting, playing the role of an on-site interaction in the mean-field approximation, leading to an additional spin-dependent contribution to the energy. Without the loss of generality, we assume that the starting spin alignment is FM, i.e. spin-up states are occupied, whereas spin-down states are empty.  The second term in Eq.~(\ref{eq:Ham_model_2}) describes an overlap between the impurity orbitals, quantified by the direct hopping $t$.

The interaction of impurity orbitals with the substrate is described through the hybridization parameter $V_h$:
 \begin{eqnarray}
\mathcal {\hat{H}}_{pd}  =  V_h \sum_{\sigma, i,j } ( \hat{d}^{\dagger}_{1 \sigma} \hat{c}_{i \sigma} + \hat{c}^{\dagger}_{i \sigma} \hat{d}_{1 \sigma}  + \hat{d}^{\dagger}_{2 \sigma} \hat{c}_{j \sigma} + \hat{c}^{\dagger}_{j \sigma} \hat{d}_{2 \sigma}),
\label{eq:Ham_model_3}
\end{eqnarray} 
where the second sum runs over three neighboring phosphorus atoms, as it is shown in Fig.~\ref{fig:BANDS}(a). In the equations above, $\hat{c}^{ \dagger}_{i \sigma}, \hat{c}_{i \sigma}$ and $\hat{d}^{\dagger}_{i \sigma}, \hat{d}_{i \sigma}$ are creation and annihilation operators for phosphorus and impurities electrons with spin $\sigma = \pm \frac{1}{2}$, respectively. In what follows, we also use the notation for spin projections $\frac{1}{2} \equiv \uparrow$ and $ -\frac{1}2 \equiv \downarrow$.  We note that here we deal with an effective model, meaning that a direct mapping of the first-principles results onto the effective model is not straightforward. The corresponding model parameters should be considered as empirical.

The band structure of our model is schematically shown in Fig.~\ref{fig:BANDS}(b). In the absence of impurity-substrate hybridization ($V_h = 0$), impurity states are represented by four energy levels.  An energy gap between the occupied and unoccupied levels is controlled by the spin splitting $\Delta$, whereas a small energy difference between each pair of levels (2$t$) originates from the direct hopping term in Eq.~(\ref{eq:Ham_model_2}). The energy of impurity states relative to the phosphorus bands structure is given by an offset energy $E_0$. 
The inclusion of hybridization leads to a finite bandwidth of impurity levels, and slightly modifies the band structure of phosphorene.

Having determined the Hamiltonian, we can estimate the exchange interaction between the impurity states. To this end, we make use of the magnetic force theorem~\cite{Local_force}:
\begin{eqnarray}
J_{12} = - \frac{1}{2 \pi S^2 }  \int_{-\infty}^{E_F} d \epsilon \Im   ( \Delta G^{\downarrow}_{12} (\epsilon) \Delta  G^{\uparrow}_{21} (\epsilon)), 
\label{eq:Local_force}
\end{eqnarray}
where $E_F$ is the Fermi energy chosen to keep constant the number of electrons, and $S=1/2$ is the spin of impurity orbitals. The one-particle Green's function $G^{\sigma}_{12} (\epsilon)$ between impurity states defined as
\begin{eqnarray}
G^{\sigma}_{12} (\epsilon) = \frac{1}{N_{k}} \sum_{\mathbf{k}, m}     \frac{e^{m}_{1 \sigma }(\mathbf{k})  e^{m}_{2  \sigma}(\mathbf{k}) } {\epsilon - E^{m}_{\sigma} (\mathbf{k})}.
\label{eq:Green_function}
\end{eqnarray}
In this equation  $e^{m}_{i \sigma }(\mathbf{k})$  stands for the $i$-th component of the $m$-th eigenvector, and $E^{m}_{\sigma} (\mathbf{k})$ is the corresponding eigenvalue of Hamiltonian [Eq. (\ref{eq:Ham_model})]. The summation runs over $N_{k}$ points of the Brillouin zone, chosen to be sufficient to reach numerical convergence of exchange integrals within 0.1 meV.
To reduce the number of free parameters, the spin splitting was fixed to $\Delta$ = 1 eV, which only influences the absolute value of exchange coupling $J_{12}$, not its sign. The other parameters, i.e., $E_{0}$, $t$ and $V_h$ were varied during the calculations.

\begin{figure}[tbp]
\includegraphics[width=0.50\textwidth]{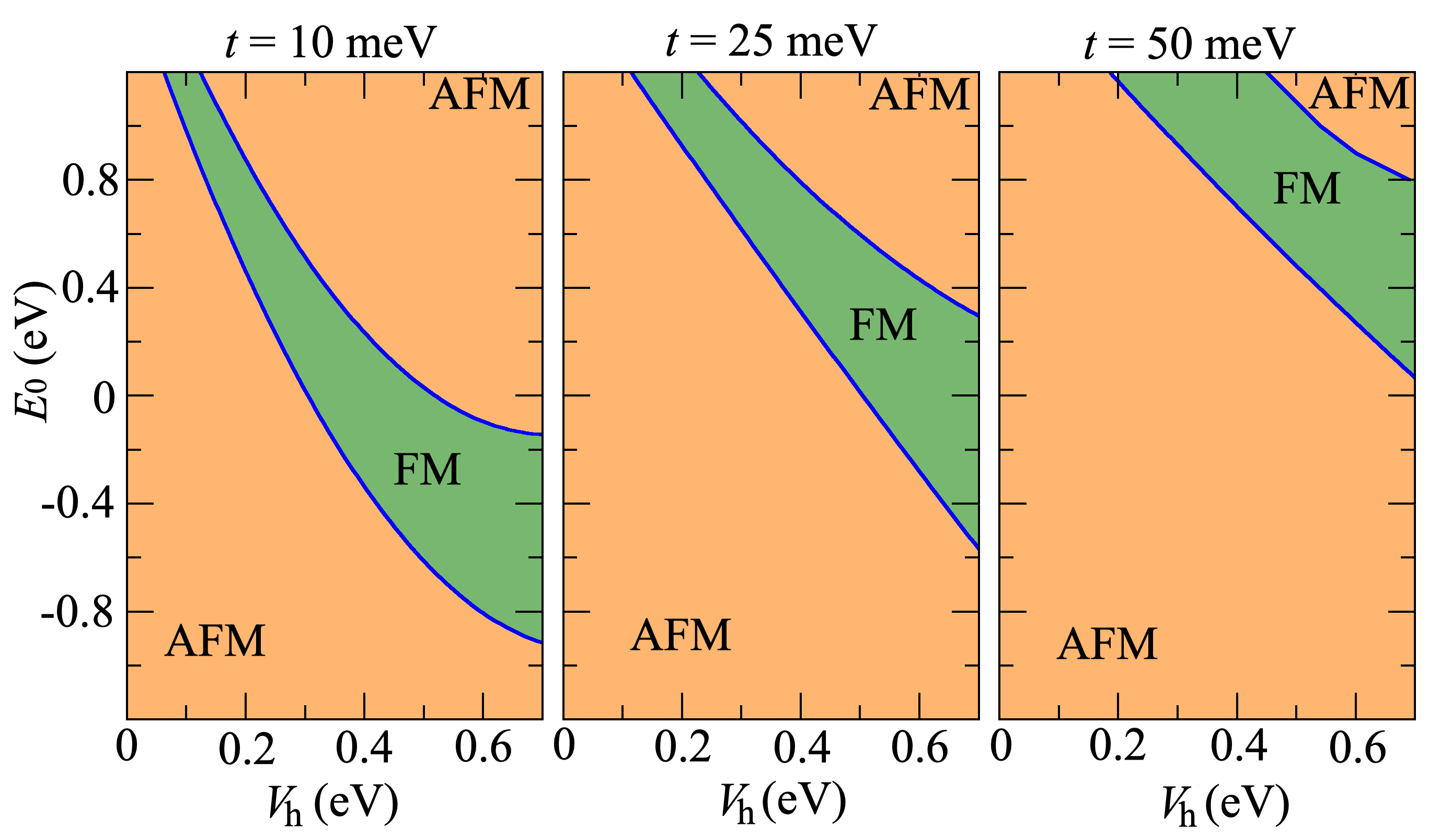}
\caption{ Phase diagram obtained from the effective two-impurity model [Eq.~(\ref{eq:Ham_model})] for three different hopping parameters $t$. The spin splitting is set to $\Delta$ = 1 eV in all calculations.}
\label{fig:Phase_diagram}
\end{figure}

The calculated phase diagram in the coordinates $E_0$ vs. $V_h$ for three different hopping parameters $t$ is shown in Fig.~\ref{fig:Phase_diagram}.  In the absence of hybridization (or if $V_h \ll t$), Eq.~(\ref{eq:Local_force}) gives an ordinary AFM superexchange coupling $J_{12} =4t^2/ \Delta$. Introduction of hybridization $V$ modifies the exchange interaction and, depending on the orbital energy $E_0$, can stabilize the FM state. If the direct hopping term $t$ is further increases, the AFM interaction becomes dominant again, suppressing the FM state, as can be seen from the phase diagram. We note, however, that a more complex substrate-mediated AFM mechanism is realized in the regime $V_h \sim t$.

Let us now make a comparative analysis of our results obtained from the effective model and DFT calculations presented in Table~\ref{tab:short_range}. Our model demonstrates that the interaction between cobalt adatoms is controlled by the direct hopping as well as by the hybridization with the substrate.  Depending on their relative magnitude, both FM and AFM coupling can be realized, as it is shown in Fig.~\ref{fig:Phase_diagram}.
  
The distance between cobalt adatoms in HH and LL states is very close to each other, meaning that the effective hopping parameters $t$ is similar in both cases. On the contrary, adatoms in the low-spin state are significantly closer to the substrate (1.01 \AA) compared to adatoms in the high-spin state (1.37 \AA).  It is natural to assume that a smaller adatom-substrate distance corresponds to a stronger hybridization, i.e. to a larger parameter $V_h$,   which also was justified by our DOS analysis in Sec~\ref{sec:Results} B. At the same time, given that FM coupling becomes more favorable for larger $V_h$, FM interaction between adatoms in the low-spin state can be attributed to an enhanced hybridization with the substrate. On the other hand, the lack of hybridization in the high-spin state of cobalt adatoms, favors AFM coupling governed by the conventional superexchange mechanism.

The interaction between adatoms in different states (HL case) has obviously a competing FM and AFM character, resulting in a quantitative reduction of the exchange interactions, as it one can see from Table \ref{tab:short_range}. Therefore, our effective model provides a qualitative explanation of the exchange interaction switching mechanism observed in the armchair direction.

\section{\label{sec:discuss}Discussion}
\begin{figure}[!tbp]
\includegraphics[width=0.48\textwidth]{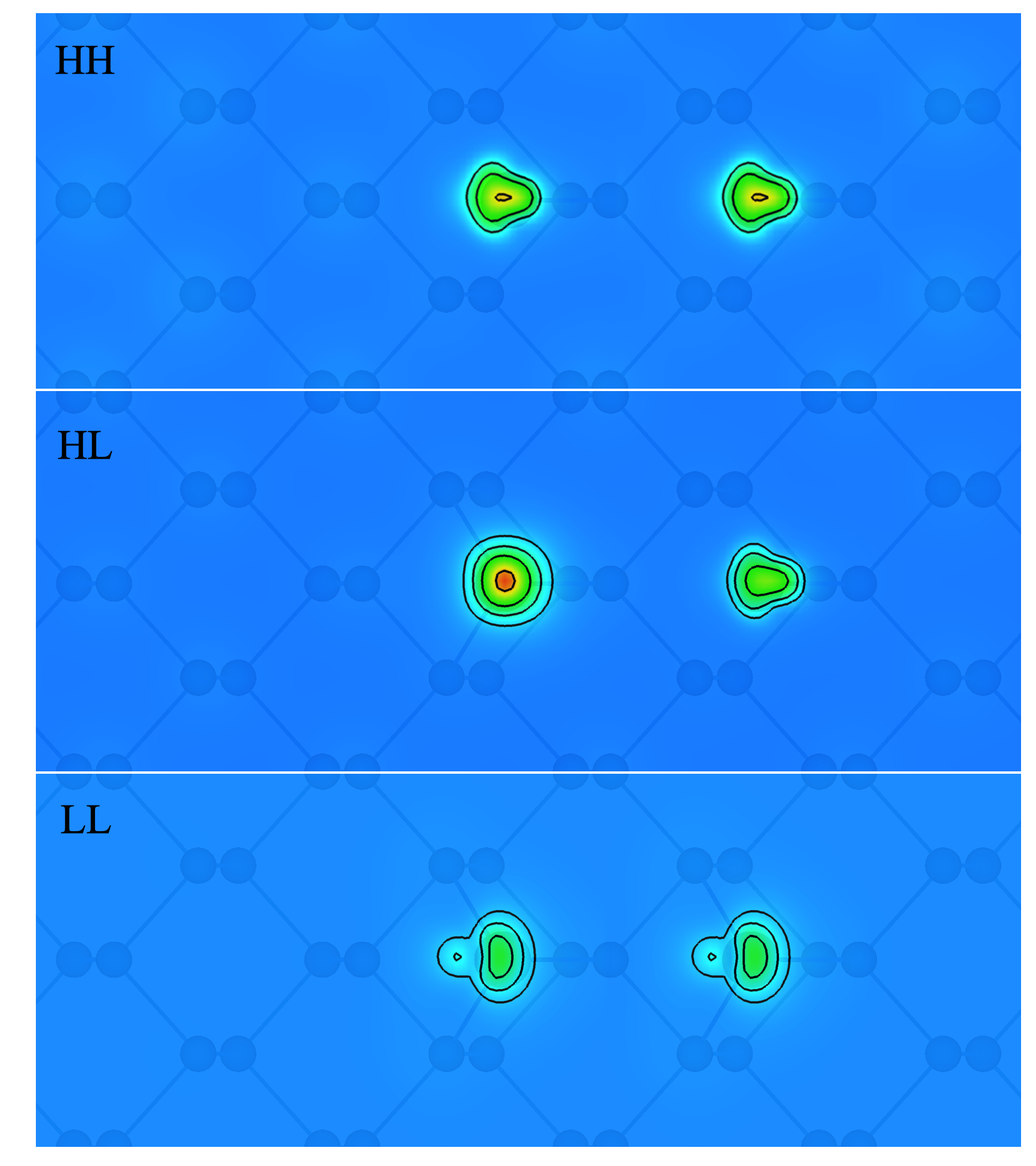}
\caption{Charge density distribution of cobalt dimer on BP calculated at the edge of the phosphorus valence band shown for the distance 1.6 \AA \ above the substrate. Different panels correspond to different spin configurations of cobalt atoms: high-high (HH), high-low (HL), and low-low (LL). The atomic configuration corresponds to the closest position of adatoms oriented along the armchair direction. Visualisation is made by {\sc vesta} software~\cite{vesta}.}
\label{fig:microscopy}
\end{figure}

We now discuss the possibility of experimental verification of the predicted exchange interaction switching. An indirect manifestation of the interaction sign can be already identified from non-spin-polarized STM topography spectra, reflecting peculiarities of charge density distribution around individual atoms. Fig.~\ref{fig:microscopy} depicts DFT charge density averaged over valence states for a pair of cobalt atoms in different spin states. The difference between various combinations of spin states can be clearly seen. While the low-spin state is characterized by a strongly localized charge density, it appears to be essentially different for the high-spin state, demonstrating a more diffusive character. Apart from varying degree of localization, the shape of charge densities corresponding to different spin states is also different. The obtained charge density of the HH and LL configurations is similar to that of single Co atoms in the high-spin and low-spin states, reported in Ref.~\onlinecite{Khajetoorians}. In contrast, the charge density of the HL configuration is notably different from the HH and LL cases.
  
Another experimental approach for probing magnetic interactions is inelastic electron tunnel spectroscopy (IETS)~\cite{Heinrich, Wiesendanger_1, Wiesendanger_2}.  In addition to the integrated characteristics (Fig.~\ref{fig:microscopy}), which can be measured by conventional STM techniques,  IETS can directly resolve the strength of magnetic interaction between surface adatoms.  The bias voltage assists in tunnelling electrons through the adatom to a conducting surface. Consequently, whenever a new conduction channel opens with increasing bias voltage, a step in the conductance spectrum appears~\cite{Delgado, Fransson}. The magnitude of bias voltage required for opening a additional conducting channel is constituting  the corresponding magnetic interaction in the system.  

From the theory side, inelastic contribution into the tunnelling current can be written in the following form~\cite{Fernandes_Rossier, Delgado}:
\begin{eqnarray}
I(V)  = \sum_{  \substack{ M, M^\prime \\ 
a, s = \pm }}   P_{M} |\braket{M| \hat{\mathbf{S}}_{a}|M^\prime}|^{2} \frac{eV - s\Delta}{1 - e^{-s \beta(eV -s\Delta)}},
\label{eq:Conductivity}
\end{eqnarray} 
where  $\Delta = \epsilon_{M^\prime} - \epsilon_{M}$ is the difference between the energies of magnetic states $\ket{M}$  and $\ket{M^\prime}$ defined in terms of the quantum Heisenberg model [Eq.~(\ref{eq:Ham})]. Parameters $\beta$ and $V$ are the inverse temperature and bias voltage, respectively. $P_{M}$  is the  occupation number of the state $\ket{M}$, which takes 1 for the ground and 0 for excited states.  $\hat{\mathbf{S}}_{a} = \sum_i  \eta (i) \hat{S}^{i}_{a}$ is the $a$-component ($a=x,y,z$) of the total spin operator, involving $i$ atoms with spin $\hat{S}^{i}_{a}$ and weight $\eta(i)$. For simplicity, we consider the situation when STM tip is placed above one of cobalt atoms in the dimer, i.e. $\hat{\mathbf{S}}_{a} = \hat{S}_{a}$.  In Eq.~(\ref{eq:Conductivity}), the matrix elements  $|\braket{M|\hat{S}_a|M^\prime}|^{2}$ have nonzero values as conditioned by the spin selection rule for total spin $\Delta S = 0, \pm1$ and its projection $\Delta S_z = 0, \pm1$~\cite{Ternes, Fernandes_Rossier, MnCuN}. Details on the quantum energy spectrum for a dimer in different spin states is given in  Appendix~\ref{app:spin_dimers}.

\begin{figure}[!tbp]
\includegraphics[width=0.5\textwidth]{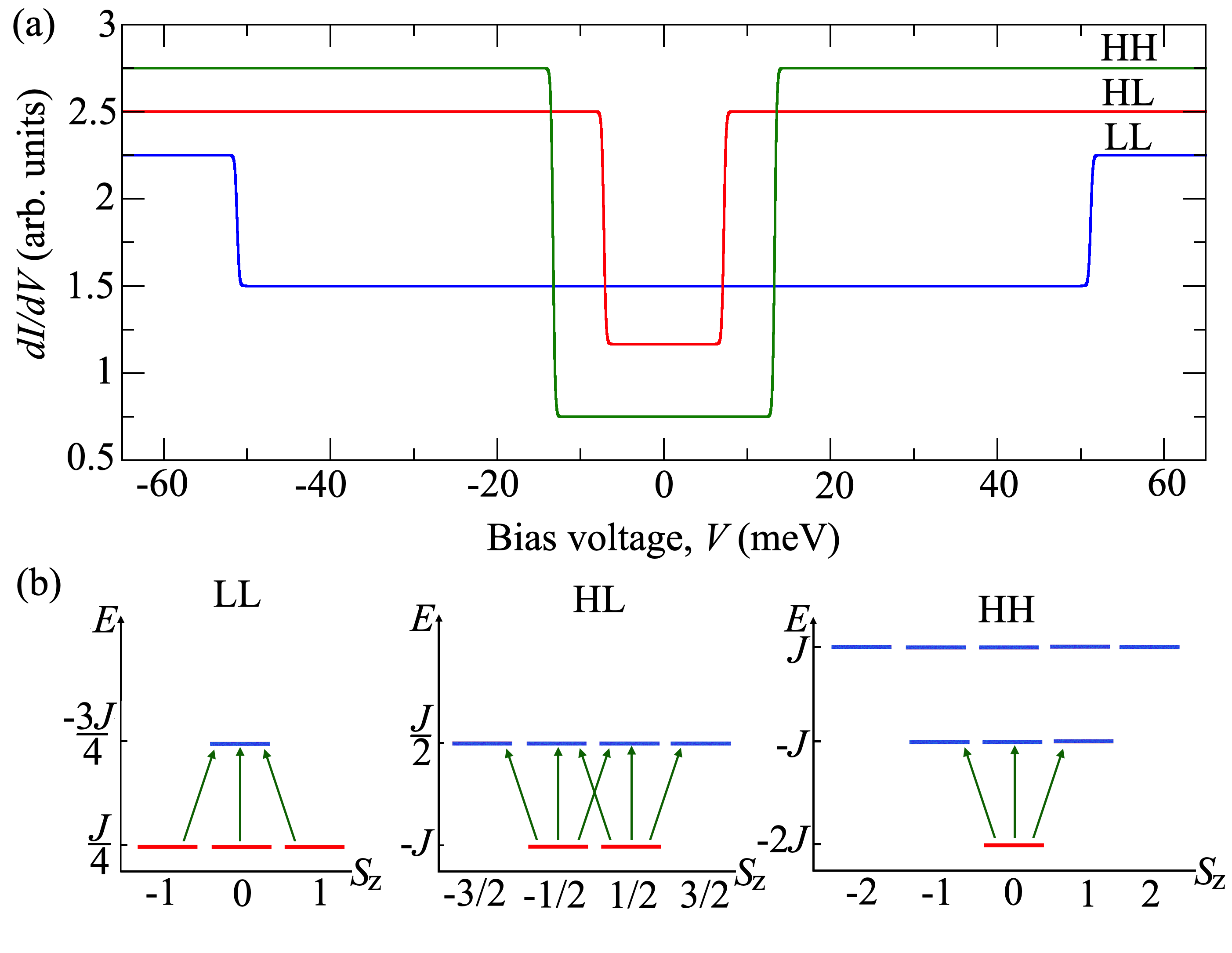}
\caption{ (a) Differential conductance  calculated for cobalt atoms, placed at nearest neighbour hollow positions along armchair direction. Here we use parameters of exchange interaction given in Table~\ref{tab:short_range}. In the simulations, temperature was fixed as $T$ = 1 K. (b) Energy levels for dimer system at LL, HL and HH spin states. The transitions between them are denoted by green arrows, contributing to the resulting differential conductance. }
\label{fig:spectroscopy}
\end{figure}

The calculated tunneling spectra obtained for the most interested case of closest cobalt adatoms oriented along the armchair direction are shown in Fig.~\ref{fig:spectroscopy}. 
The most simple case is realized for the  LL spin configuration with FM coupling ($J<0$). In this situation, the ground state is represented by a degenerate triplet state with the energy $\frac{J_{12}}{4}$. The singlet state is separated by an energy gap, whose magnitude is equal to the exchange interaction $J_{12}$. As soon as bias voltage reaches the value of $J_{12}$, a step appears in the $dI/dV$ spectrum [Fig.~\ref{fig:spectroscopy}(a)], manifesting itself a triplet-singlet transition. 

For  HL spin configuration, total spin of the system can be either $S = 1/2$ or $3/2$, according to the momentum sum rules: $|S_1 - S_2| \leq S \leq S_1 + S_2$. The ground state of AFM-coupled ($J>0$) HL dimer is thus given by a doubly degenerate state $S = 1/2$ with the energy $E = -J_{12}$. This state is separated from quadruplet excited states ($S =3/2$) by an energy gap  $\frac{3}{2} J_{12}$.  The transition between the states takes place only if the selection rule $\Delta S_z = 0, \pm1$ is satisfied. The allowed transitions are shown in Fig.~\ref{fig:spectroscopy}(b).
Next, in case of AFM-coupled HH spin state there are three multiplets: the ground state with total spin $S = 0$ and two excited states ($S=1$ and $2$). The corresponding energies read $-2 J_{12}$ (singlet), $-J_{12}$ (triplet), and $J_{12}$ (quintuplet). The allowed transitions correspond to singlet-triplet excitations, occurring at the bias voltage $J_{12}$, which can be seen from Fig.~\ref{fig:spectroscopy}(a). Transitions to the highest excited state is suppressed by the selection rule $\Delta S = 0, \pm1$, leading to zero matrix element in Eq. (\ref{eq:Conductivity}). Therefore, one can see that three different spin states of cobalt dimer on BP exhibit distinct conductance spectra, and can be resolved experimentally.

Finally, it is worth mentioning that IETS measurements require a conductive substrate. In case of BP, we do not expect considerable technical issues because BP is known to be intrinsically $p$-doped, allowing for tunneling conductance \cite{Kiraly2017}. Alternatively, the substrate can be heavily doped by alkali metals, as it is done, e.g., in Ref.~\onlinecite{Kiraly2019}. These manipulations might lead to additional RKKY-like contribution to the exchange arising from the coupling of adatom spins with conduction electrons of the substrate. However, considering an essentially long-range character of indirect interactions, we do not expect that they considerably affect the short-range switching behavior predicted in our work. Particularly, the conductance spectra restored in Fig.~\ref{fig:spectroscopy} should be reliable for the experimentally required regimes.

\section{\label{sec:conc}Conclusion}
We have reported on a prospective mechanism for magnetic interaction control between adatoms by means of field-assisted repopulation of orbital states. By considering the example of cobalt adatoms deposited on phosphorene, we have theoretically demonstrated that an FM-AFM reversible switching of exchange interactions between adatoms can be realized upon changing the bias voltage. 
Our model analysis reveals that hybridization with the substrate, being dependent on the spin state, plays a key role in the switching of magnetic interactions. Simulations of STM spectra performed on the basis of quantum Heisenberg model suggests that the predicted effect can be observed experimentally. Our prediction is not limited to Co/BP system, but can be generalized to other transition metal atoms deposited on insulating substrates.

\begin{acknowledgements}
This work was supported by the Russian Science Foundation, Grant No. 17-72-20041. D.I.B. acknowledges the Russian Federation Presidential scholarship for providing travel support to visit the Radboud University. The authors would like to thank Alexander Khajetoorians and Brian Kiraly for stimulating discussions regarding experimental realization of our idea, and for their useful comments on our paper.
\end{acknowledgements}

\appendix

\section{\label{app:ISO}Exchange interactions from first principles}
In Table~\ref{tab:Exchange}, we present a summary of our first-principles data on exchange interactions $J_{12}$ between two cobalt atoms in different spin states calculated along the armchair and zigzag directions, as well as the corresponding total magnetic moments $M$ and optimized interatomic distances $R$.

\begin{table}[tbp]
\centering
\caption [Bset]{Summary of exchange interactions $J_{12}$ (in meV), and total  magnetic moments $M$ (in $\mu_B$) calculated for different distances (in \AA) between cobalt adatoms deposited on phosphorene along the \emph{armchair} and \emph{zigzag} directions. The combination of different spin states is denoted as high-high (HH), high-low (HL), and low-low (LL).}
\begin{ruledtabular}
\begin {tabular}{c|cccccccc}
 &  &  \multicolumn{3}{c}{ \emph{armchair}} &   &  \multicolumn{3}{c}{ \emph{zigzag}} \\
 \cline{3-5} \cline{7-9} 
 &    &  $R$ &  $J_{12}$  &  $M $  &     &  $R$ &  $J_{12}$  &  $M $  \\
  \hline
\multirow{ 3}{*}{1}  &  HH &  4.34  &  13.3  &  4.20  &      &   2.82  &  -6.1     &   4.20  \\
                                &   HL &   4.47  &   4.8  &  3.00  &      &   3.25  &  -25.3   &  2.32  \\
                                &   LL  &   4.37  & -51.2  &  2.00  &     &    3.22  & -145.5   &  2.00  \\
  \hline
\multirow{ 3}{*}{2}  &  HH &  5.33 &   11.4  &  4.65   &      &   5.01  &   1.0     &   4.85  \\
                                &   HL &   5.44  &   6.5  &  3.00  &      &   5.03  &  -2.6   &   2.95  \\
                                &   LL  &  5.53  &    4.5  &  2.00   &     &   5.05  &   5.4    &  2.00  \\
  \hline
\multirow{ 3}{*}{3}  &  HH &  8.75  &   9.3   &  4.80   &      &   6.63  &   2.0     &   4.98  \\
                                 &   HL &  8.65  &  10.8  &  3.00  &      &    6.64  &   1.5     &   3.00 \\
                                 &   LL  &  8.75  &  -3.3  &  2.00   &     &    6.63   & -20.9    &  2.00  \\
  \hline
\multirow{ 3}{*}{4}  &  HH &  9.59  &   6.4   &  5.00   &      &   8.31  &    0.6     &   5.00  \\
                                 &   HL &  9.70  &  1.6  &  3.10  &      &    8.33   &   0.5     &   3.05 \\  
                                 &   LL  &  9.80   &  2.2  &   2.00   &     &    8.32   &  -4.1     &  2.00  \\    
                                   \hline
\multirow{ 3}{*}{5}  &  HH &  13.12  &   2.9   &  5.00   &      &    9.94  &  -0.3      &   5.00  \\  
                                 &   HL &  13.02  &   3.7   &  3.20  &      &    9.95   &   0.2     &   3.10 \\    
                                 &   LL  &  13.12   &  9.4   &   2.00   &     &    9.94   &  -3.4     &  2.00  \\                         
                                 
\end {tabular}
\end{ruledtabular}
\label{tab:Exchange}
\end{table}

\section{\label{app:spin_dimers}Quantum spectrum of a spin dimer}
Table \ref{tab:Spectra} presents energy spectrum of the quantum Heisenberg Hamiltonian  [Eq.(\ref{eq:Ham})] with the corresponding eigenvectors for the following combinations of two spins: (i) $S_1 = \frac{1}{2}, S_2 = \frac{1}{2}$ (LL); (ii) $S_1 =1, S_2 = \frac{1}{2}$ (HL); and (iii) $S_1 = 1, S_2 =1$ (HH).

\begin {table}[!tbp]
\centering
\caption [Bset]{Energy spectrum of the quantum Heisenberg Hamiltonian [Eq.~(\ref{eq:Ham})] with the corresponding eigenstates obtained for three different combinations of spins $S=1/2$ and $1$. The following notation for spin projections is used: $\ket{\uparrow} = 1/2$, $\ket{\downarrow}= -1/2$, and $\ket{\Uparrow} = 1$, $\ket{\varnothing}= 0$, $\ket{\Downarrow}= -1$ for spins $S=1/2$ and $1$, respectively.    }
\begin{ruledtabular}
\setlength{\extrarowheight}{5pt}
\begin {tabular}{c|c}
\multicolumn{2}{c}{ LL}\\
   \hline
$-\frac{3}{4} J_{12}$ &  $\frac{1}{\sqrt{2}} (\ket{\uparrow \downarrow} - \ket{\downarrow \uparrow})$ \\
$\frac{1}{4} J_{12}$ & $\ket{\uparrow \uparrow}$,  $\frac{1}{\sqrt{2}} (\ket{\uparrow \downarrow} + \ket{\downarrow \uparrow})$, $\ket{\downarrow \downarrow}$ \\
   \hline
 \multicolumn{2}{c}{ HL}\\
    \hline
  $-J_{12}$ & $-\frac{2}{\sqrt{6}} \ket{\Downarrow \uparrow} + \frac{1}{\sqrt{3}} \ket{\varnothing \downarrow}$, $-\frac{1}{\sqrt{3}} \ket{\varnothing \uparrow} + \frac{2}{\sqrt{6}} \ket{\Uparrow \downarrow}$  \\
 $\frac{1}{2} J_{12}$ &  $\ket{\Downarrow \downarrow}$,  $ -\frac{1}{\sqrt{3}} \ket{\Downarrow \uparrow } - \frac{2}{\sqrt{6}} \ket{\varnothing \downarrow}$, $-\frac{2}{\sqrt{6}} \ket{\varnothing \uparrow} - \frac{1}{\sqrt{3}} \ket{\Uparrow \downarrow}$, $\ket{\Uparrow \uparrow}$  \\  
    \hline
 \multicolumn{2}{c}{ HH}\\
    \hline
    $-2J_{12}$ & $\frac{1}{\sqrt{3}} (\ket {\Downarrow \Uparrow} - \ket{\varnothing \varnothing} + \ket{\Uparrow \Downarrow} )$ \\
    $-J_{12}$   &  $\frac{1}{\sqrt{2}} ( \ket{\varnothing \Downarrow}  - \ket{\Downarrow \varnothing})$, $\frac{1}{\sqrt{2}} (\ket {\Uparrow \Downarrow} -\ket{\Downarrow \Uparrow})  $, $\frac{1}{\sqrt{2}} (\ket{\Uparrow \varnothing }  - \ket{\varnothing \Uparrow})$ \\
    \multirow{ 2}{*}{$J_{12}$}    &   $\ket{\Downarrow \Downarrow}$, $\frac{1}{\sqrt{2}} (\ket{\Downarrow \varnothing} + \ket{\varnothing \Downarrow})$,  $\frac{1}{\sqrt{6}} (\ket{\Downarrow \Uparrow}   + 2 \ket{\varnothing \varnothing}  + \ket{\Uparrow \Downarrow})$,  \\ 
                  &    $\frac{1}{\sqrt{2}} \ket{\varnothing \Uparrow} + \ket{\Uparrow \varnothing}$,  $\ket{\Uparrow \Uparrow}$ \\  

\end {tabular}
\end{ruledtabular}
\label{tab:Spectra}
\end{table}

\newpage

%

\end{document}